\documentclass[a4paper,11pt]{article}

\usepackage[english]{babel}
\usepackage{fullpage}
\usepackage{siunitx}
\usepackage[colorlinks=true, linkcolor=blue, citecolor=blue, urlcolor=blue]{hyperref}
\usepackage[utf8]{inputenc}
\usepackage{authblk}
\usepackage{graphicx}
\usepackage{float}
\usepackage[T1]{fontenc}            
\usepackage{newtxtext,newtxmath}    
\usepackage{subcaption}             
\usepackage{gensymb}                
\usepackage{booktabs}               
\usepackage{anyfontsize}
\usepackage{natbib}
\usepackage[normalem]{ulem}
\usepackage{soul}

\newcommand{\angstrom}{\textup{\AA}} 

\title{\Huge\textbf{Broadband Low-Resolution Spectrograph - {\sc SpectrumMate LR}}}

\author[1,2]{\textbf{Nguyen-Duc Nguyen\thanks{Corresponding author: ducnguyen382002@gmail.com}}}
\author[3]{\textbf{Le-Quang Thuy}}
\author[4]{\textbf{Tobias C. Hinse}}
\author[3]{\textbf{Nguyen-Van Tue}}
\author[2,5,6]{\textbf{Nguyen-Luong Quang}}

\affil[1]{University of Science and Technology of Hanoi, Department of Space and Applications, Hanoi, Vietnam}
\affil[2]{TNU Observatory, Tay Nguyen University, Buon Ma Thuot, Dak Lak, Vietnam}
\affil[3]{Quy Nhon Observatory, ExploraScience, Quy Nhon, Binh Dinh, Vietnam}
\affil[4]{University of Southern Denmark, Department of Physics, Chemistry and Pharmacy, SDU-Galaxy, Campusvej 55, 5230 Odense M, Denmark}
\affil[5]{CSMES, The American University of Paris, PL111, 2 bis, passage Landrieu, 75007, Paris, France}
\affil[6]{Université Paris-Saclay, Université Paris Cité, CEA, CNRS, AIM, 91191, Gif-sur-Yvette, France}

\date{\today}

\begin{document}
\maketitle

\begin{abstract}
This paper presents the development and application of {\sc SpectrumMate LR}, a broadband, low-resolution spectrograph tailored for use with small telescopes. SpectrumMate LR is designed to offer affordable, accessible spectroscopic capabilities for amateur astronomers and educators, inspired by the need for versatile instruments in amateur and educational settings. Utilizing a 300 $grooves/mm$ grating, 80 $mm$ collimator and objective lenses, {\sc SpectrumMate LR} is optimized to perform analyses across the visible spectrum, enabling users to classify stars by spectral type, measure stellar temperatures, and test filter transmission ranges. Tests demonstrate SpectrumMate LR’s ability to capture accurate spectral data, validating its efficacy in observing both celestial and terrestrial light sources. This instrument fills a niche for cost-effective spectroscopy, empowering a broader audience to engage in detailed observational astronomy.

\end{abstract}

\section{Introduction}

Spectroscopy is a crucial tool for understanding stars' and other celestial bodies' physical and chemical properties. By analyzing the spectrum of a light source, we can deduce essential characteristics such as temperature, chemical composition, and age, along with motion, distance, and more subtle features like magnetic fields and atmospheric conditions. For centuries, spectroscopic techniques have deepened our knowledge of the universe, revealing not only the life cycles of stars but also the conditions in distant galaxies, the composition of exoplanetary atmospheres, and the large-scale structure of the cosmos.

In recent years, advancements in spectroscopic technology have enhanced our ability to capture and analyze celestial spectra, but the high costs and complexity of traditional equipment often place it out of reach for amateur astronomers and small observatories. These barriers have limited widespread engagement in astronomical spectroscopy, despite the immense value that even lower-resolution data can provide in educational settings and citizen science projects.

In Vietnam, the Nha Trang Observatory is equipped with a medium-resolution eShel II spectrograph $(R = 10000\ \textnormal{at } \lambda=6563\ \angstrom\textnormal{, which is around the H-alpha line})$, while Quy Nhon Observatory\footnote{\url{https://astro.explorascience.vn/}} (QNO) hosts the low-resolution \textsc{SpectrumMate} spectrograph $(R = 2666\ \textnormal{at } \lambda=5500\ \angstrom)$. Although these instruments are effective for high-precision spectral observations, neither can capture the entire visible spectrum in a single image. This limitation poses challenges for applications such as chemical composition analysis and spectral classification, which often require a broader spectral coverage. To address this gap, we have developed the broadband spectrograph presented in this paper, named \textsc{SpectrumMate LR}.

{\sc SpectrumMate LR} is a broadband, low-resolution spectrograph specifically optimised for small telescopes. Designed with accessibility and affordability in mind, the {\sc SpectrumMate LR} brings spectroscopic capabilities to smaller observatories, amateur astronomers, and educational institutions. Its user-friendly setup and compatibility with standard telescope mounts make it an ideal tool for individuals and organisations eager to participate in spectroscopic observations and data collection. With the ability to observe a wide range of wavelengths at a low resolution, {\sc SpectrumMate LR} allows users to perform meaningful analyses of stellar spectra, tracking changes over time and comparing characteristics across various types of stars and other celestial objects.

{\sc SpectrumMate LR}’s design emphasises both usability and data quality. Its broadband sensitivity spans the visible spectrum, making it suitable for observations of many different kinds of objects, from bright stars to nebulae. The low-resolution capability allows for a quick and reliable assessment of spectral features such as absorption and emission lines, essential for identifying key elements within a star or nebula. Although the instrument is limited in its ability to resolve very fine spectral details, its broadband, low-resolution approach provides a sufficient level of information for a wide variety of analyses, enabling users to identify prominent spectral features and obtain general characteristics of the observed objects.

By making spectroscopic analysis more accessible, {\sc SpectrumMate LR} opens new pathways for amateur astronomers and educators to contribute to scientific research. This tool offers a hands-on way to engage with the same fundamental techniques used by professional astronomers, facilitating a broader understanding of the universe and inspiring new generations to explore astrophysics through direct observation. The {\sc SpectrumMate LR} aims to democratise spectroscopic data collection, empowering users worldwide to participate in astronomy at a deeper level.

\section{{\sc SpectrumMate LR} design}
\subsection{Requirements}
In contrast to the original {\sc SpectrumMate}, which has a spectral coverage of $368\ \angstrom$ in one image \citep{phan2024}, {\sc SpectrumMate LR} is engineered to capture a broader range of wavelengths, from $3400\ \angstrom$ to $7000\ \angstrom$. This enhanced wavelength range enables the observation of a wide spectrum of celestial features, from the hydrogen and helium lines typical in stellar spectra to more complex molecular signatures found in nebulae and other diffuse objects. The increased bandwidth and optimised resolution make the {\sc SpectrumMate LR} suitable for capturing both broad spectral features and essential details of emission and absorption lines. 

{\sc SpectrumMate LR} design requirements prioritise accessibility, user-friendliness, and compatibility with small telescopes, addressing the need for affordable and efficient spectroscopic tools in amateur astronomy and educational settings. There are several key requirements for {\sc SpectrumMate LR} so that it will be able to make high-quality, broadband spectroscopic analysis accessible to a wider audience, enabling educational institutions and amateur astronomers to explore the universe in greater depth.

Firstly, broad spectral coverage. Covering the visible spectrum allows for versatile analysis across different types of celestial bodies. This range captures visible light and near-infrared or near-ultraviolet spectra, depending on the user’s observational goals, making the device suitable for a wide range of astronomical applications.

Secondly, moderate resolving power. With a resolving power of $R=990$ at $5500 \angstrom$, the middle of the visible wavelength range, {\sc SpectrumMate LR} achieves a balance between detail and accessibility. While it does not reach the high resolutions needed for detailed spectral line measurements in professional research, this level of resolution is ideal for detecting major spectral lines, such as hydrogen alpha (H$\alpha$) and beta (H$\beta$) lines, sodium, magnesium, and other elements, providing valuable data for amateur and educational use.

Thirdly, high light sensitivity. The device's sensitivity must be optimised to perform well with small-aperture telescopes, typically 127 to 254 $mm$ (5 to 10 $inches$) in diameter and focal ratios from about f/5 to f/10, which are common in amateur astronomy. This ensures that \textsc{SpectrumMate LR} can capture sufficient light from fainter objects, such as stars of magnitude 10 or 12, without prohibitively long exposure times, while maintaining a manageable field of view for the spectrograph's setup.

Next, ease of installation and calibration. Designed for compatibility with standard mounts and small telescopes, the {\sc SpectrumMate LR} should allow for straightforward setup and calibration procedures. Features like an integrated calibration lamp or easy access to external calibration sources are essential to ensure that users of all skill levels can quickly calibrate and maintain accuracy in their observations.

Finally, high portability and durability. Given its intended use in amateur astronomy and educational outreach, the device should not be heavier than 2 $kg$ and durable, allowing for transport to various observation sites. This portability expands the tool’s utility, allowing users to bring it to remote locations with minimal setup, maximising the flexibility of observation opportunities.

\subsection{Theoretical optical designs}
\begin{figure}[H]
	\centering
	\includegraphics[width=\linewidth]{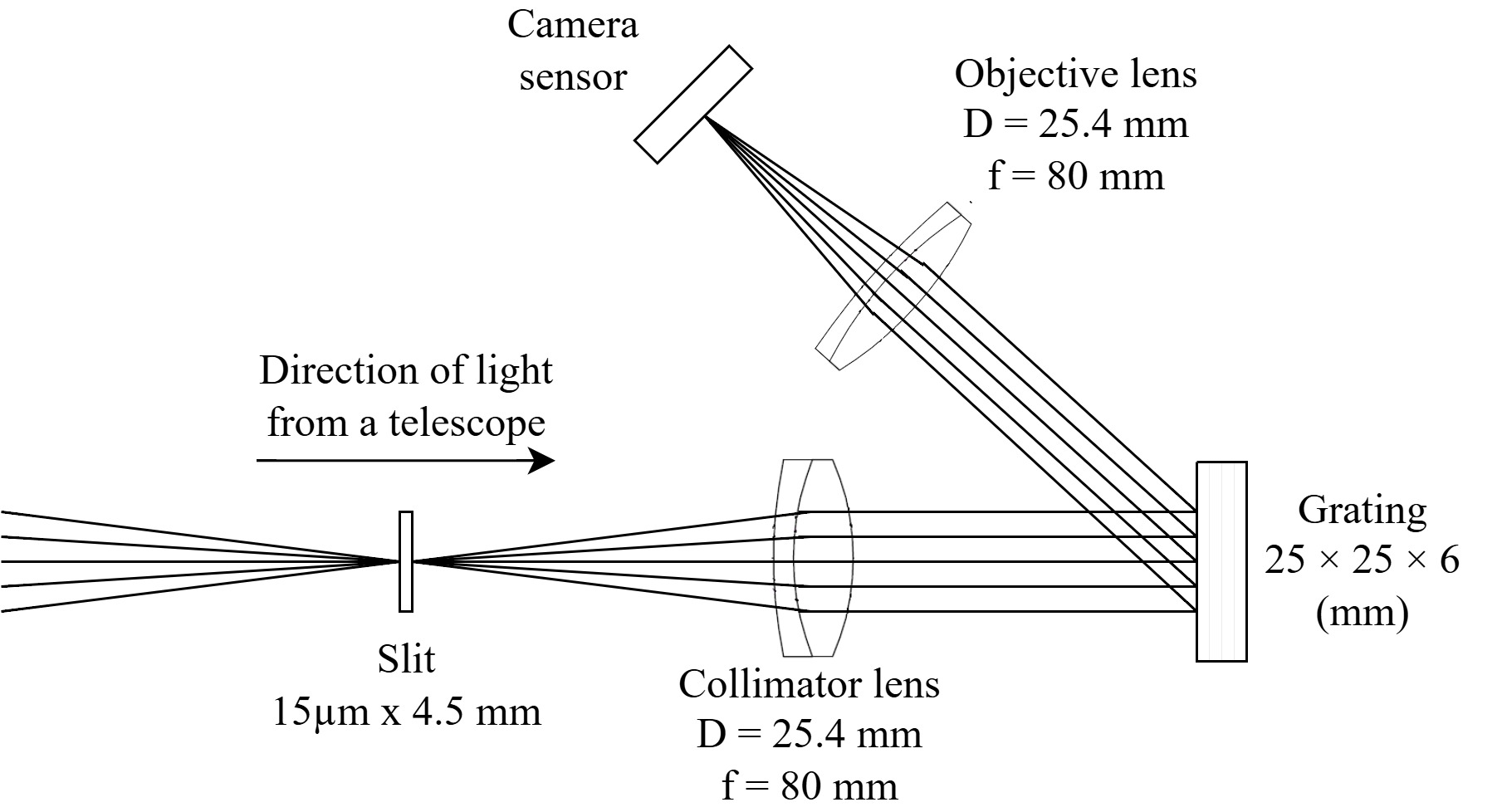}
	\caption{\textsc{SpectrumMate LR} optical system. Compared to \textsc{SpectrumMate}, the lenses and the grating are the main differences.}
    \label{fig:11111}
\end{figure}

The theoretical background provided in the work of \citet{builpara, buil2024} and \citet{phan2024} offers a basis for the optical system of \textsc{SpectrumMate LR} (\autoref{fig:11111}), determining optimal spectral range, resolution, and other design parameters to enhance performance for small-telescope compatibility. The two mentioned works also cover telescope aperture compatibility and sensitivity of a spectrograph, which are not covered in this paper.

\autoref{fig:11111} shows the optical system of \textsc{SpectrumMate LR}. The collimator lens (OP0176) and objective lenses (OP0182) utilised in this spectrometer are both achromatic doublets, specifically designed to minimise chromatic aberration, thereby significantly reducing colour distortion across the entire visible wavelength range. The objective lens is also visible optimised, ensuring a high transmission in this range.

Following the theoretical framework established, we calculated the \textsc{SpectrumMate} LR parameters to meet the specific requirements for broadband, low-resolution spectroscopic applications. The parameters marked with an asterisk in \autoref{table:parapara} are not a spectrograph component and may vary depending on the instrument used.

\begin{table}[H]
	\centering
	\caption{Parameters used for calculation. The calculated parameters for {\sc SpectrumMate LR} provide details on the achievable wavelength range, spectral resolution, and light sensitivity.}
        \label{table:parapara}
	\begin{tabular}{ll}
		\toprule
		Parameter & {\sc SpectrumMate LR} \\
		\midrule
		Collimator lens: focal length $f_1$ and diameter $D_1$ & $80\ mm;\ 25.4\ mm$ \\
		$F_C=f_1/D_1$ & $1.97$ \\
		Objective lens: focal length $f_2$ and diameter $D_2$ & $80\ mm;\ 25.4\ mm$ \\
		$F_O=f_2/D_2$ & $1.97$ \\
		The total angle between the incident ray and the diffracted ray $G$ & $34\degree$ \\
		Distance from the slit to the collimator lens & $80\ mm$ \\
		Distance from the objective lens to the camera & $80\ mm$ \\
		Grating groove density $m$ * & $300\ \textnormal{grooves}/mm$ \\
		Grating size * & $25\times25\times6\ mm$ \\
            Slit width $w$ * & $15\ \mu m$ \\
		Telescope’s principal mirror diameter $D$ and focal length $f$ * & $610\ mm;\ 3962\ mm$ \\
		Sensor dimensions * & $5496\times3672\ pixel$ \\
		Pixel size $p$ * & $2.4\times2.4\ \mu m$ \\
		\bottomrule
	\end{tabular}
\end{table}

\subsubsection{Vignetting check at the collimator}
The aperture ratio $F_\#$ of the telescope will be \begin{equation}F_\#=\dfrac{f}{D}=\dfrac{3962}{610}=6.5\end{equation}

The collimator lens will be large enough to pass all the light from the telescope if $F_C<F_\#$. Since $F_C = 1.97 < 6.5$, this condition is satisfied.

\subsubsection{Beam diameter}

We denote the diameter of the beam exiting the collimator as $d_1$. In this case, we find \begin{equation}
d_1=D\left(\frac{f_1}{f}\right)=610\times\left(\frac{80}{3962}\right)=12.32\, mm
\end{equation}
assuring the correct dimension of the collimator lens diameter.

\subsubsection{Angles of incidence and diffraction}
The spacing between the grooves in the grating determines the angle at which the different wavelengths of light are diffracted. The grating equation, given by 

\begin{equation}
k\lambda=d(\sin{\alpha}+\sin{\beta})
\end{equation}
relates the wavelength of the light $\lambda$, the distance between the grooves $d$, the angle of incidence $\alpha$, the angle of diffraction $\beta$ (both in the unit of $degree$), and the order of the diffraction $k$. It is sometimes convenient to write the equation as 

\begin{equation}
mk\lambda=\sin{\alpha}+\sin{\beta}
\end{equation}
where $m=1/d$ is the groove density (grooves per millimetre).

\begin{figure}[H]
	\centering
	\includegraphics[width=\linewidth]{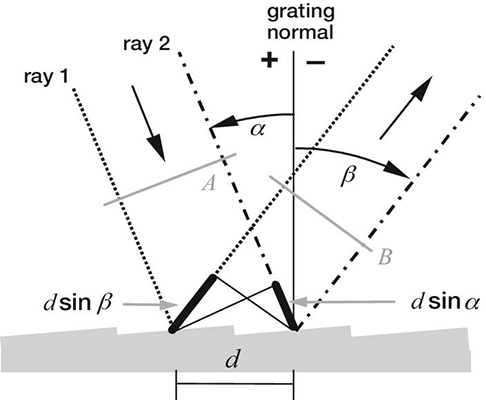}
	\caption{Grating geometry by \citet{palmerdiff}.}
    \label{gratinggeometry}
\end{figure}

The equations presented are applicable only when the incident and diffracted rays are situated in a plane that is perpendicular to the grooves at the centre of the grating, as shown in \autoref{gratinggeometry}. This condition, called classical or in-plane diffraction, encompasses most grating systems \citep{palmerdiff}.

For the design of {\sc SpectrumMate LR}, the light of wavelength $\lambda=5500\, \angstrom$ is aimed to be located at the centre of the imaging sensor. We have chosen the centre of the visible spectrum (green) as the reference wavelength.

We have $G = \alpha-\beta\ $, replace $\beta$ with $G$, the grating equation can be written as

\begin{equation}
\sin\alpha+\sin(\alpha-G)=mk\lambda
\end{equation}

For maximum S/N performance, working with the first order $k=\pm1$ (high light intensity) spectrum is generally recommended. Here we only work on the first order $k=1$. Solving for $\alpha$, the above equation becomes

\begin{equation}
\begin{split}
    \alpha &= \arcsin\left(\frac{mk\lambda}{2\cos(G/2)}\right) + \frac{G}{2} \\
           &= \arcsin\left(\frac{1\times300\times5500\times10^{-7}}{2\cos(34/2)}\right) + \frac{34}{2} \\
           &= 21.95
\end{split}
\end{equation}

From this we find

\begin{equation}
\beta=\alpha-G=21.95-34=-12.05
\end{equation}

Note that angles of opposite signs mean the incident ray and the diffracted ray lie on opposite sides of the normal to the grating.

\subsubsection{Minimum grating dimension}
For the required minimum grating dimension $L$, we have 

\begin{equation}
    L=\frac{d_1}{\cos\alpha}=\frac{12.32}{\cos21.95\degree}=13.28\ mm
\end{equation}

Since $L<25\ mm$, the chosen holographic diffraction grating is large enough for no light loss to occur.

\subsubsection{Spectral dispersion on sensor}
Let $\rho$ be the degree of dispersion in $\angstrom/pixel$ on the camera sensor. To optimise data storage, we use $2 \times 2$ binning, which groups four neighbouring pixels into one "superpixel" that has the size doubled as the original pixel. Therefore, the effective pixel size \( p \) is doubled in dispersion calculations.

The dispersion is given by
\begin{equation}
    \rho = 10^7 \times \frac{p\times\cos \beta}{m\times f_2}
\end{equation}
which yields
\begin{equation}\label{spectraldispersion}
    \rho = 10^7 \times \frac{2.4 \times 10^{-3}\times{2} \times \cos(-12.05^\circ)}{300 \times 80} = 1.96\ \angstrom/pixel.
\end{equation}

\subsubsection{Diameter $d_2$ of the diffracted rays}
The beam diameter \( d_2 \) of diffracted rays reflected off from the grating is given by
\begin{equation}
d_2 = \frac{\cos \beta}{\cos \alpha}\times\frac{f_1}{F_\#} + \frac{X\times p \times N}{f_2}
\end{equation}
where \( X \) is the distance from the grating to the objective lens, and \( N \) is the number of pixels across the sensor in the dispersion direction. Due to mechanical constraints, \( X \) is set to \( 80 \, mm \). For the chosen sensor (Sony IMX183CLK-J), \( N = 5496 \) pixels.

Although \(2 \times 2\) binning is used for data collection, we utilise the pixel size $p=2.4\ \mu m$ pixel count \(N = 5496\) in this calculation. This is because the beam diameter \(d_2\) depends on the actual physical parameter of the sensor, rather than on the effective value after binning. Binning does not alter the physical dimensions of the sensor, which directly affects the beam spread.

Substituting values, we find
\begin{equation}
d_2 = \frac{\cos(-12.05^\circ)}{\cos(21.95^\circ)} \times \frac{80}{6.5} + \frac{80 \times 2.4 \times 10^{-3} \times 5496}{80} = 26.16 \, mm.
\end{equation}

For the total absence of vignetting of the sensor, the condition \( F_O < f_2 / d_2 \) must be fulfilled. We calculate \( 1.97 < 3.06 \) for our setup, thus satisfying the required condition.

\subsubsection{Spectral range coverage}

The extreme wavelength limits from one side of the sensor to the other are given by the relation
\begin{equation}
\label{spectralrange}
\lambda_{1,2} = \lambda \mp \frac{N \rho}{2} = 5500 \mp \frac{\dfrac{5496}{2} \times 1.96}{2}
\end{equation}

Here, \(N\) represents the full number of pixels (5496) across the sensor width. Although we employ \(2 \times 2\) binning to save data, which reduces the effective spatial resolution, the physical width of the sensor remains unchanged. Therefore, we use the original pixel count (5496) to calculate the spectral range. Binning only affects the sampling resolution and not the physical span of the sensor, which determines the wavelength range coverage.

Substituting values, we find

\begin{equation}
\lambda_1 = 2806.96\ \angstrom, \quad \lambda_2 = 8193.04\ \angstrom
\end{equation}

Thus, {\sc SpectrumMate LR} in this configuration covers a spectral range of \(5386.08\ \angstrom\) with spectral dispersion of $1.96\ \angstrom/pixel$.

\subsubsection{Spectral resolution and resolving power}
The spectral resolving power $R$ is defined as
\begin{equation}
R = \frac{\lambda}{\Delta\lambda}
\end{equation}
where $\Delta\lambda$ is the full-width half maximum (FWHM) of the spectral peak being investigated. The higher the value of $R$, the higher the spectral resolution. {\sc SpectrumMate LR} is of low resolution with a value of $R = 524$, which tells us that {\sc SpectrumMate LR} allows one to see details of $5500/524=10.50\ \angstrom$ in the range near $5500\ \angstrom$.

\subsection{Mechanical Design}
Before initiating the 3D printing process, a comprehensive mechanical design was developed using Catia V5\footnote{\url{https://www.3ds.com/products/catia/catia-v5}}. We adopted a modular design approach to facilitate ease of assembly, maintenance, and potential future upgrades. The design process involved detailed simulations verifying component tolerances, ensuring proper optical alignment, and minimising light leakage. Specific design solutions were implemented to guarantee robust interfaces between critical components, such as maintaining an optimal distance between the collimator lens and the slit and ensuring precise alignment of the objective tube with the camera system.

After completing these design and validation steps, the finalised 3D models were converted into STL files for printing on an Ender 3 printer. Due to its ease of use, rapid prototyping capability, and cost-effectiveness, PLA filament was chosen as the primary material for \textsc{SpectrumMate LR}'s enclosure and parts, rather than traditional aluminium milled components. Although aluminium offers superior mechanical stability and thermal resistance, PLA was deemed sufficient for our application under typical operating conditions. Moreover, we carefully evaluated the potential for PLA deformation under varying telescope positions and ambient temperatures. Our tests, including thermal and alignment verifications, indicated that any deformation of the PLA parts is minimal and does not adversely affect the optical alignment or overall performance of the system. Routine calibration and maintenance procedures are also in place to mitigate any minor misalignment that could arise over time.

\begin{figure}[H]
    \centering
    \includegraphics[width=\linewidth]{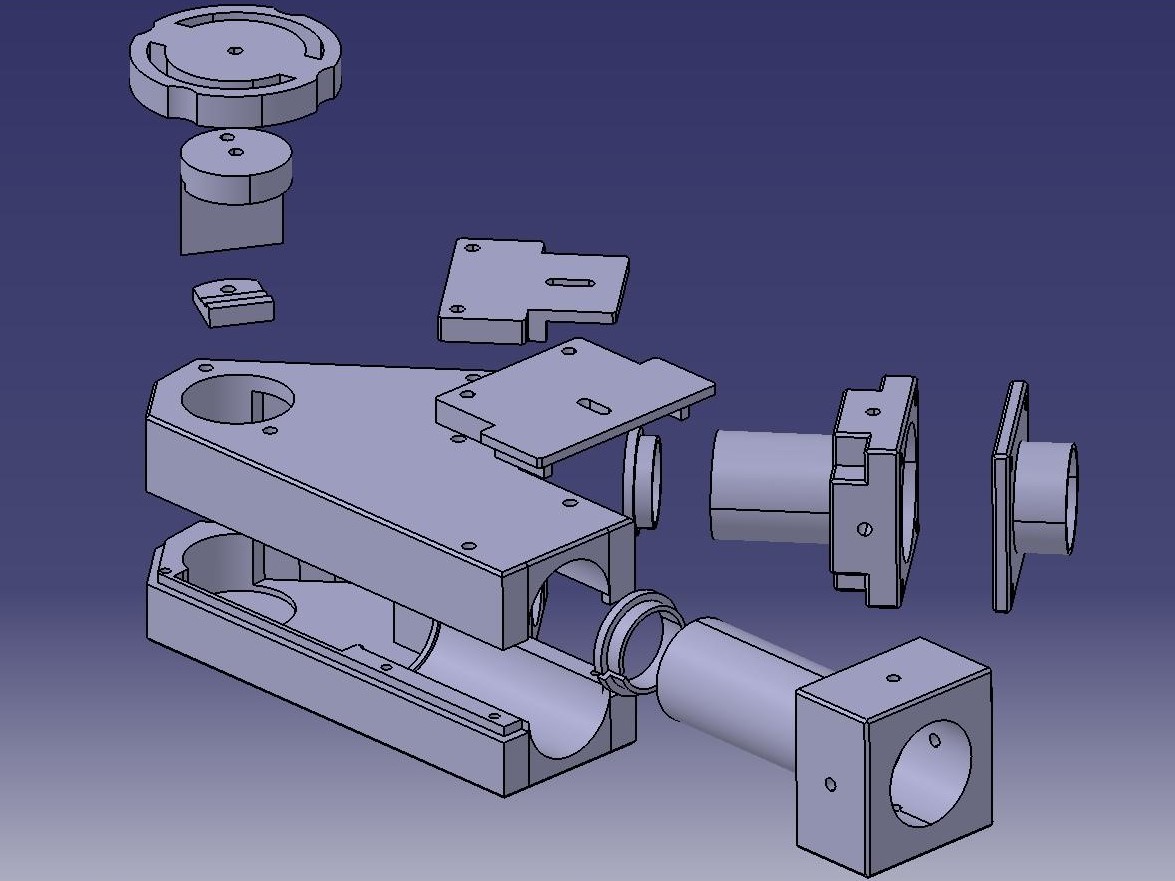}
    \caption{3D model of {\sc SpectrumMate LR}.}
    \label{fig:3D model}
\end{figure}

Images of the components can be seen in \autoref{fig:3D model}.

For the collimator lens and its interface, when the lens and slit were installed, the distance between them was maintained at approximately $50\ \textnormal{mm}$, ensuring a robust setup. The objective tube was tested with the ASI183 Pro camera and a focuser, directed at a distant object to simulate parallel rays, successfully achieving focus at infinity. These validations confirm that \textsc{SpectrumMate LR} functions as intended.

Basic tools such as pliers, screwdrivers, M3/M4 Allen keys, and various M3/M4 screws, nuts, threads, and washers were used to assemble the printed parts. Once assembled, ensuring that light entered and passed through the system without obstruction or leakage was the highest priority. A laser test confirmed proper alignment and revealed no unintended light blocking within the optical path. When the slit was blocked, no stray light was detected in the camera image, thereby affirming that \textsc{SpectrumMate LR} effectively prevents light leakage.

\begin{figure}[H]
    \centering
    \includegraphics[width=\linewidth]{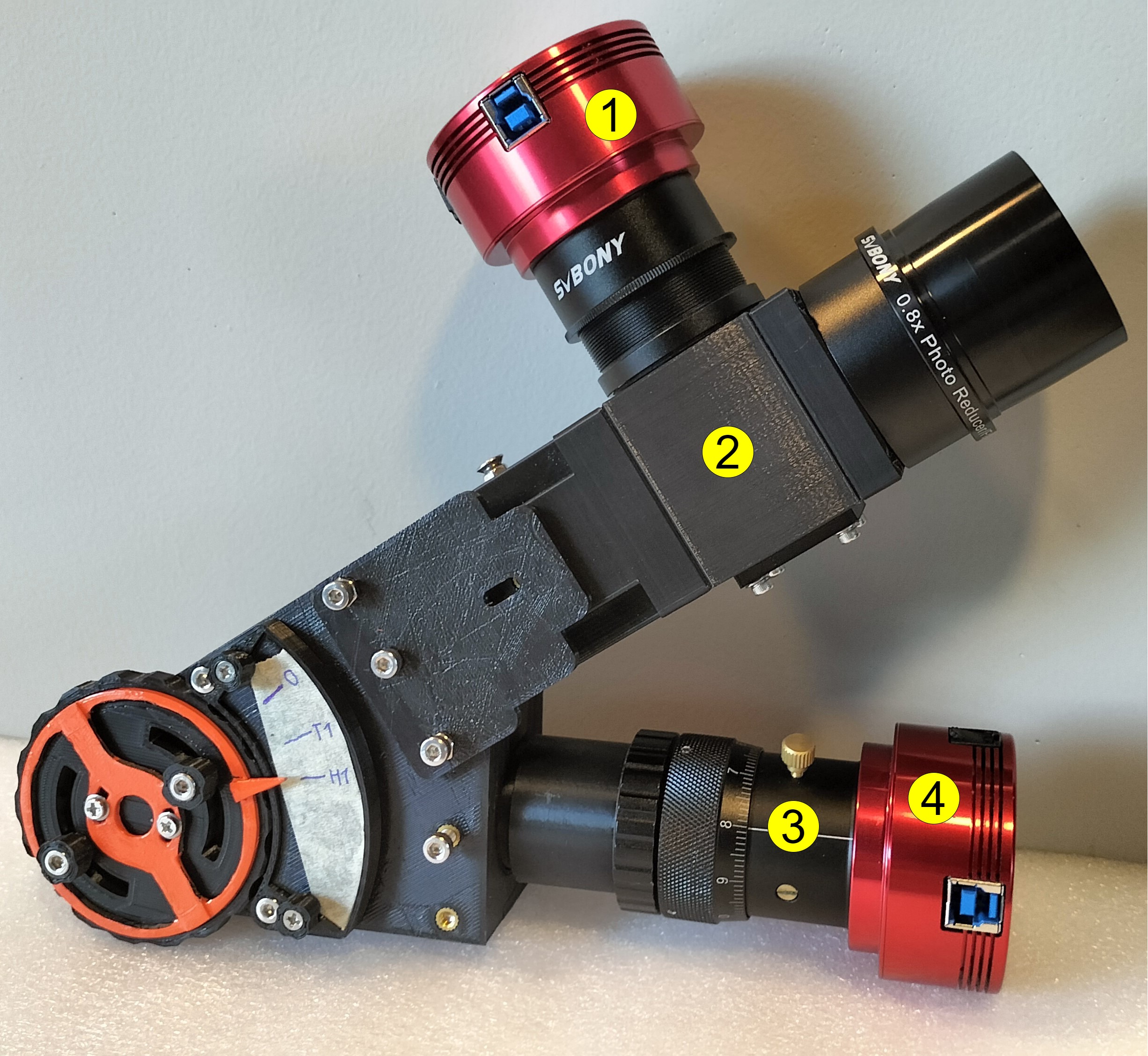}
    \caption{{\sc SpectrumMate LR} equipped with accessories. (1) is a guiding camera, that allows the user to see what the slit is pointing to by looking at the guider (2). (3) is a helical focuser, used for precisely changing the focus of the main camera (4), which captures the spectra.}
    \label{fig:Untitled Diagram}
\end{figure}

\autoref{fig:Untitled Diagram} shows an image of \textsc{SpectrumMate LR} with several accessories attached. The image caption provides a short description of the components.

\autoref{fig:Screenshot 2023-08-29 234832} below illustrates the field of view (FOV) of the spectrograph as observed through the guider system, clearly depicting the central slit (indicated by the black line) that selectively admits the light from the target object.

\begin{figure}[H]
    \centering
    \includegraphics[width=\linewidth]{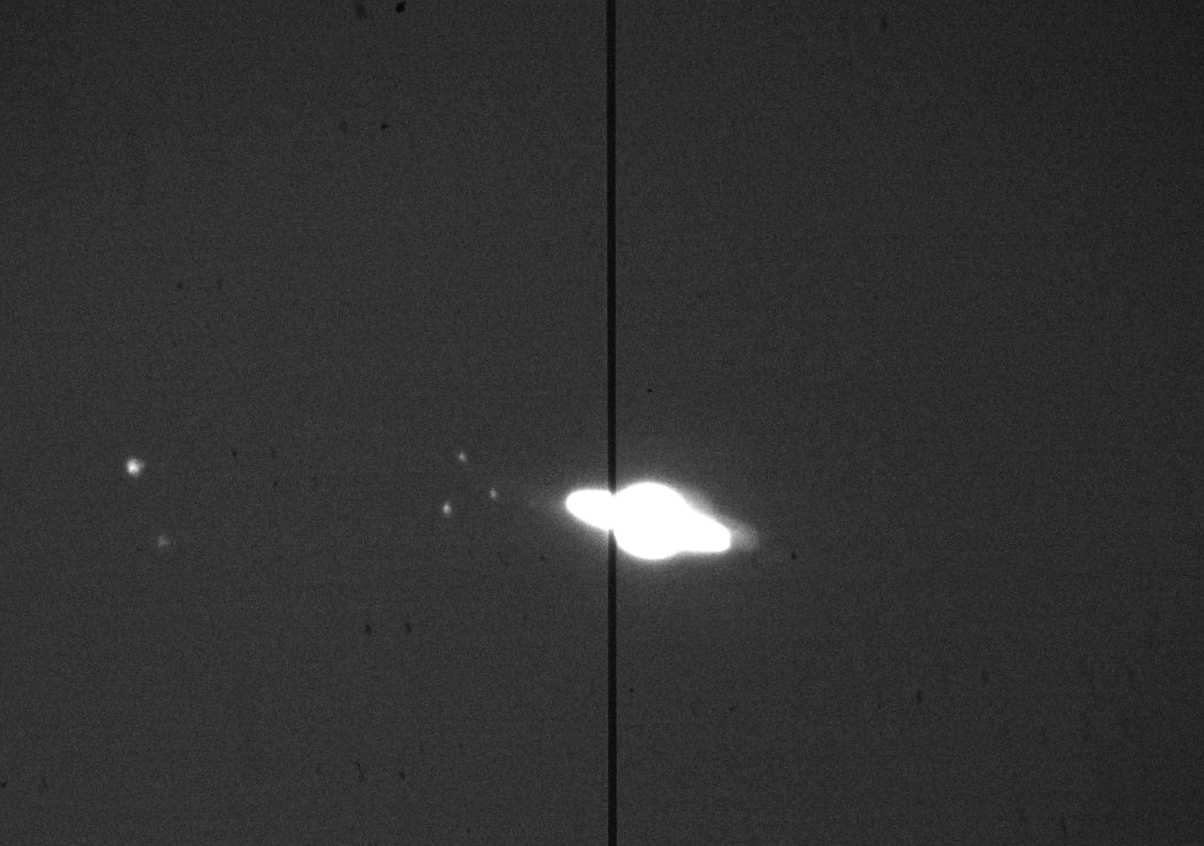}
    \caption{FOV of the spectrograph through the guider. The black line in the middle is the slit. The light of the object that the slit is pointing to will go inside the spectrograph. In this case, light from the ring of Saturn.}
    \label{fig:Screenshot 2023-08-29 234832}
\end{figure}

\section{Observations}
This section describes the observational capabilities and tests conducted with {\sc SpectrumMate LR} to evaluate its performance. These observations aimed to classify stars by spectral type, measure the temperature of different light sources, and validate the spectral transmission properties of various optical filters. Through these tests, we verified the resolution, accuracy, and reliability of {\sc SpectrumMate LR} for amateur and educational spectroscopy.

\subsection{Raw spectra obtained by \textsc{SpectrumMate LR}}
\begin{figure}[H]
    \centering
    \includegraphics[width=\linewidth]{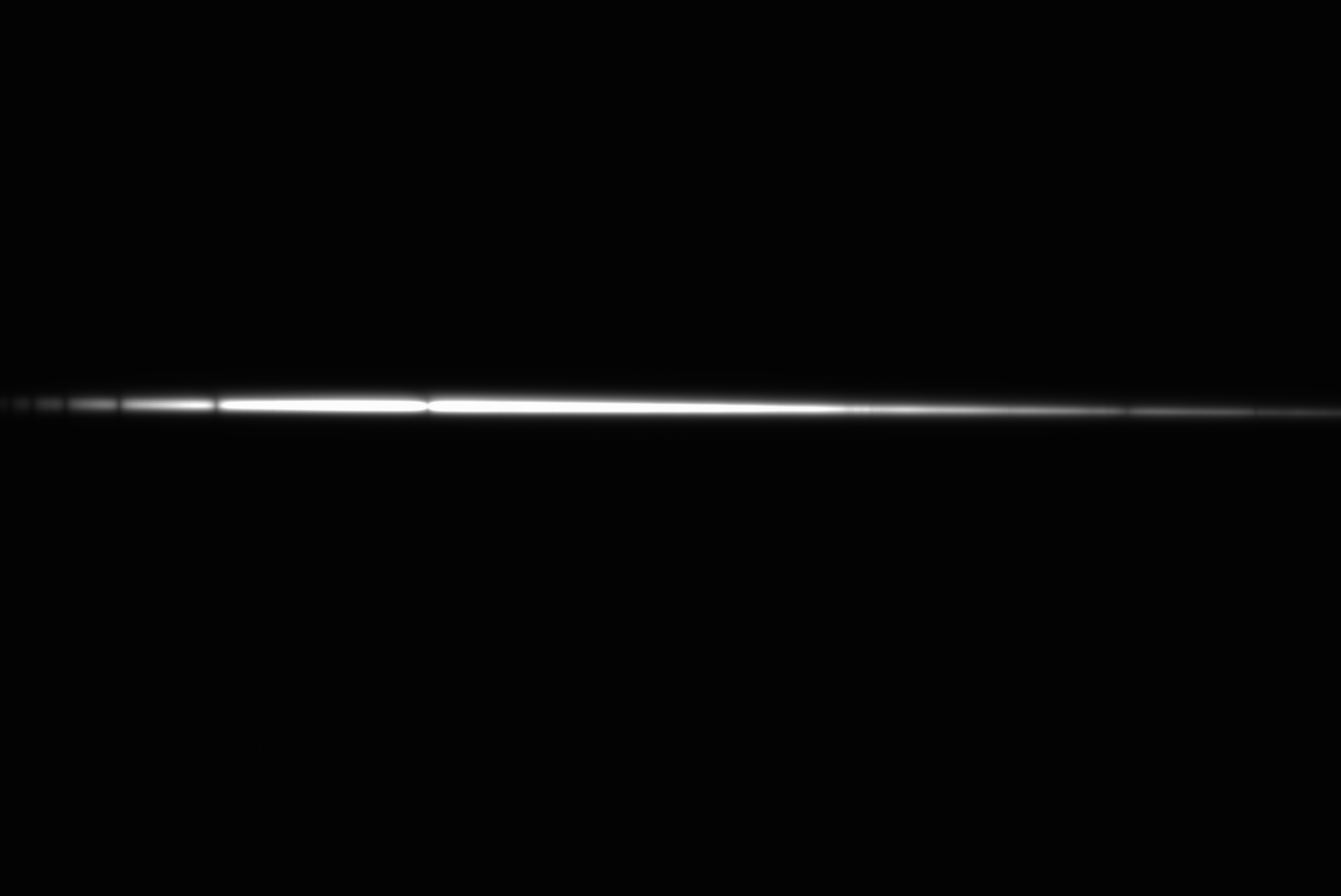}
    \caption{Raw spectrum of Vega. Absorption lines can be seen.}
    \label{fig:Vega_Object_5s_20231111_120358}
\end{figure}

\autoref{fig:Vega_Object_5s_20231111_120358} shows the raw spectrum of Vega, captured by \textsc{SpectrumMate LR}

\begin{figure}[H]
    \centering
    \includegraphics[width=\linewidth]{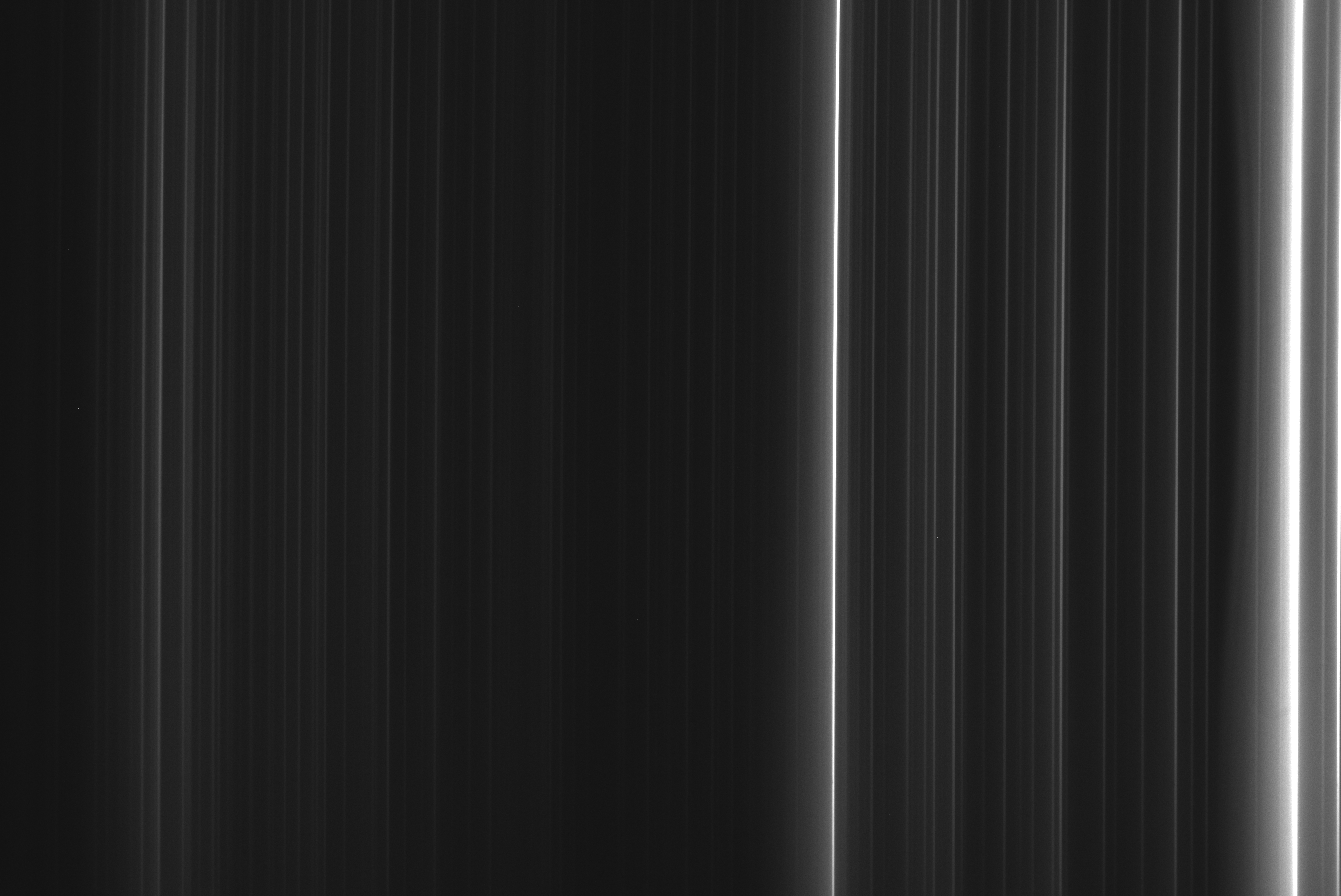}
    \caption{Calibration spectrum. It was also used for measuring the spectral resolution of \textsc{SpectrumMate}.}
    \label{fig:Calibration_180s_20231111_180535-4}
\end{figure}

To increase the quality of the spectrum, we performed the necessary calibration on the image (bias/dark subtraction, flat fielding, and stacking). For spectral calibrating and normalisation, we used the calibration module from Shelyak\footnote{\url{https://www.shelyak.com/produit/pf0037-module-detalonnage-alpy/?lang=en}}. They included a reference spectrum for the module, and the wavelength of emission lines from this module (\autoref{fig:Calibration_180s_20231111_180535-4}) can be automatically recognised if one uses Demetra\footnote{\url{https://www.shelyak.com/software/demetra/?lang=en}} software. Demetra also assists users in calibrating the images, which reduces the weight of work dramatically.

\subsection{Star Classification Based on Spectral Type}
To test the capabilities of {\sc SpectrumMate LR}, we observed and classified stars across various spectral classes, including O, B, A, F, G, K, M, and Wolf-Rayet types. These classifications are based on spectral features characteristic of each class, such as specific absorption lines, which correlate with stellar temperature and composition.

We obtained the spectra using the CDK600 telescope at QNO. The mount's accurate tracking system allowed us to take spectra of even low-brightness objects. This telescope has supported the first detection of exoplanet transit in Vietnam, by \citet{Tue-Nguyen-Van2023}, and was the main instrument used in SAGI Summer School 2023, detail is shown in the paper written by \cite{Quang-NGUYEN-LUONG2023}.

A set of representative stars from each spectral class was selected and observed to capture their spectra. In the present work, we focus on the relative shape and features of each star’s spectrum, rather than its absolute flux scale. Since the final spectra are normalised, the plotted $y$-axis units are arbitrary but comparable from star to star.

Below are examples of stars in each spectral class captured with {\sc SpectrumMate LR}, demonstrating the instrument’s ability to resolve and distinguish these spectral features. For a complete set of spectra, please refer to the supplementary materials available at \url{https://doi.org/10.5281/zenodo.14848606}.

\begin{figure}[H]
    \centering
    \includegraphics[width=\linewidth]{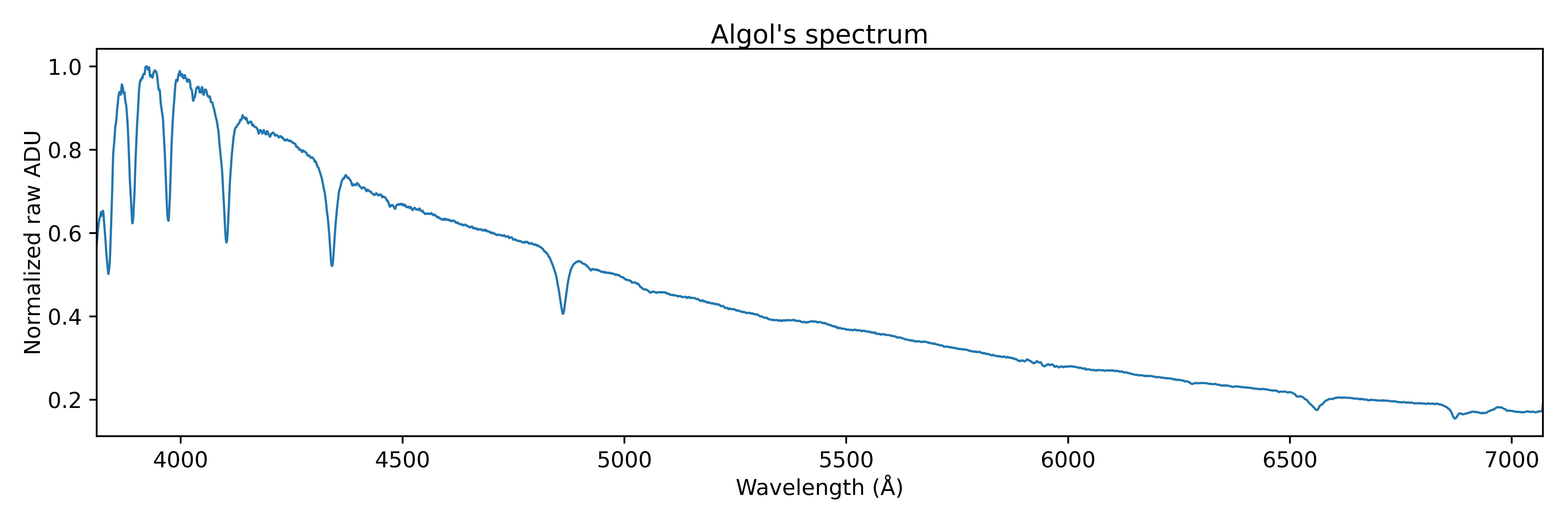}
    \caption{Algol (HD 19356) - B8V.}
\end{figure}

\begin{figure}[H]
    \centering
    \includegraphics[width=\linewidth]{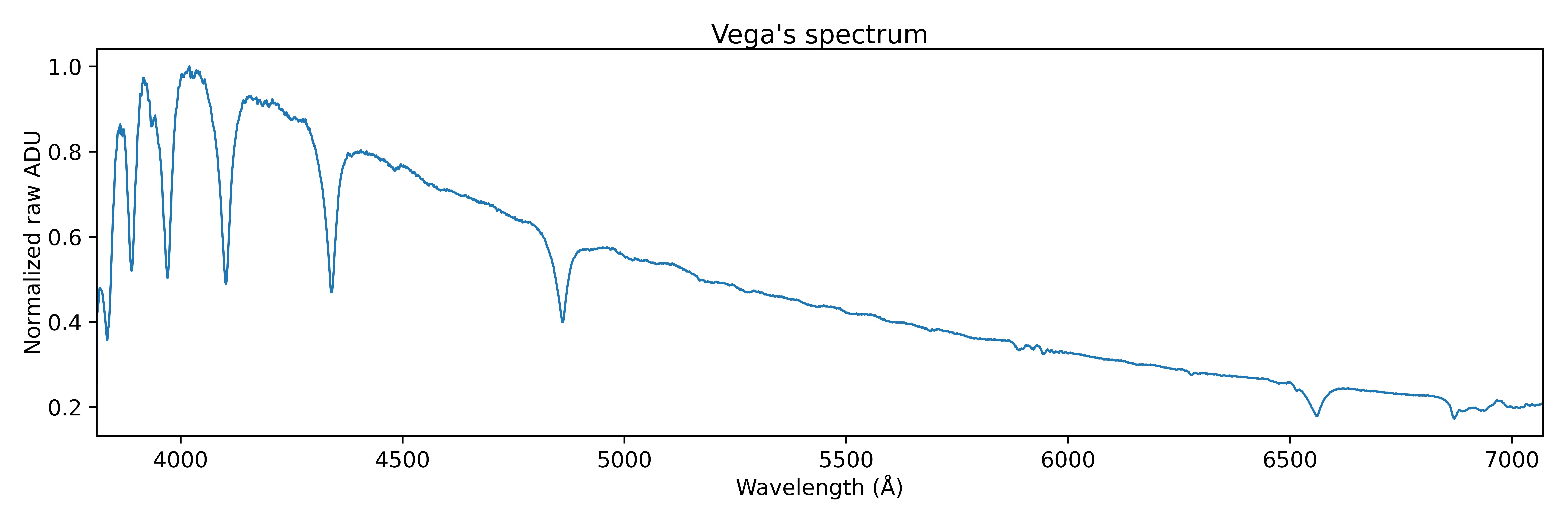}
    \caption{Vega (HD 172167) - A0V.}
\end{figure}

\begin{figure}[H]
    \centering
    \includegraphics[height=0.8\textheight]{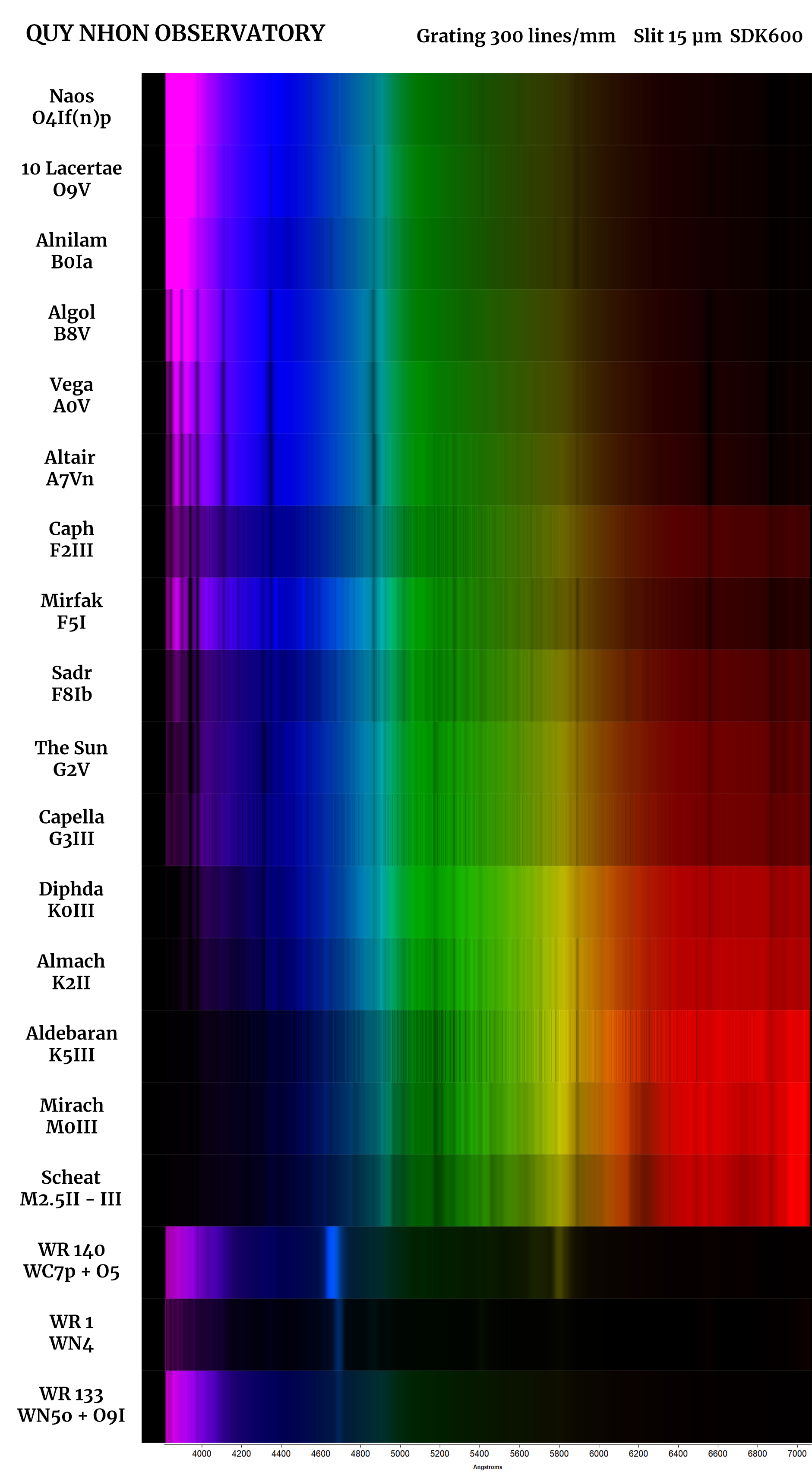}
    \caption{Star classification based on spectral types. The hotter the star, the shorter the peak wavelength at which it emits light. For the high-resolution version, visit \url{https://doi.org/10.5281/zenodo.14230992}.}
    \label{fig:Untitled-1}
\end{figure}

Demetra also generated colour spectra for each star, making it easier to analyse and categorise based on the visualised spectral lines. In \autoref{fig:Untitled-1}, we put the spectra in order of descending star temperature, except for the last three.

Based on the calibrated spectrum, the spectral dispersion and spectral range coverage of \textsc{SpectrumMate LR} were measured using Demetra. As shown in \autoref{comparison}, the spectral range coverage and resolving power are lower than theoretical calculations. Real-world performance often falls below theoretical calculations due to minor mechanical misalignment, imperfect focus, manufacturing variations in optics and gratings, and small calibration errors. Even slight deviations in component angles, lens coatings, or sensor alignment can shift the actual spectral range on the detector and reduce resolution. Thus, small, cumulative discrepancies between design assumptions and physical realities lead to lower-than-expected coverage and resolving power.

\begin{table}[H]
    \centering
    \caption{Parameters of {\sc SpectrumMate LR}.}
    \label{comparison}
    \begin{tabular}{lcc}
        \toprule
        & Theory & Real \\
        \midrule
        Spectral dispersion $\rho\ (\angstrom/\text{pixel})$ & $1.96$ & $1.18$ \\
        Spectral range coverage ($\angstrom$) & $5386.08$ & $3260.15$ \\
        Spectral resolution ($\angstrom$) & $10.50$ & $13.64$ \\
        Resolving power $R$ (at 5500 $\angstrom$) & $524$ & $403$ \\
        \bottomrule
    \end{tabular}
\end{table}

\subsection{Temperature Measurement of Light Sources}
\begin{figure}[H]
    \centering
    \includegraphics[width=\textwidth]{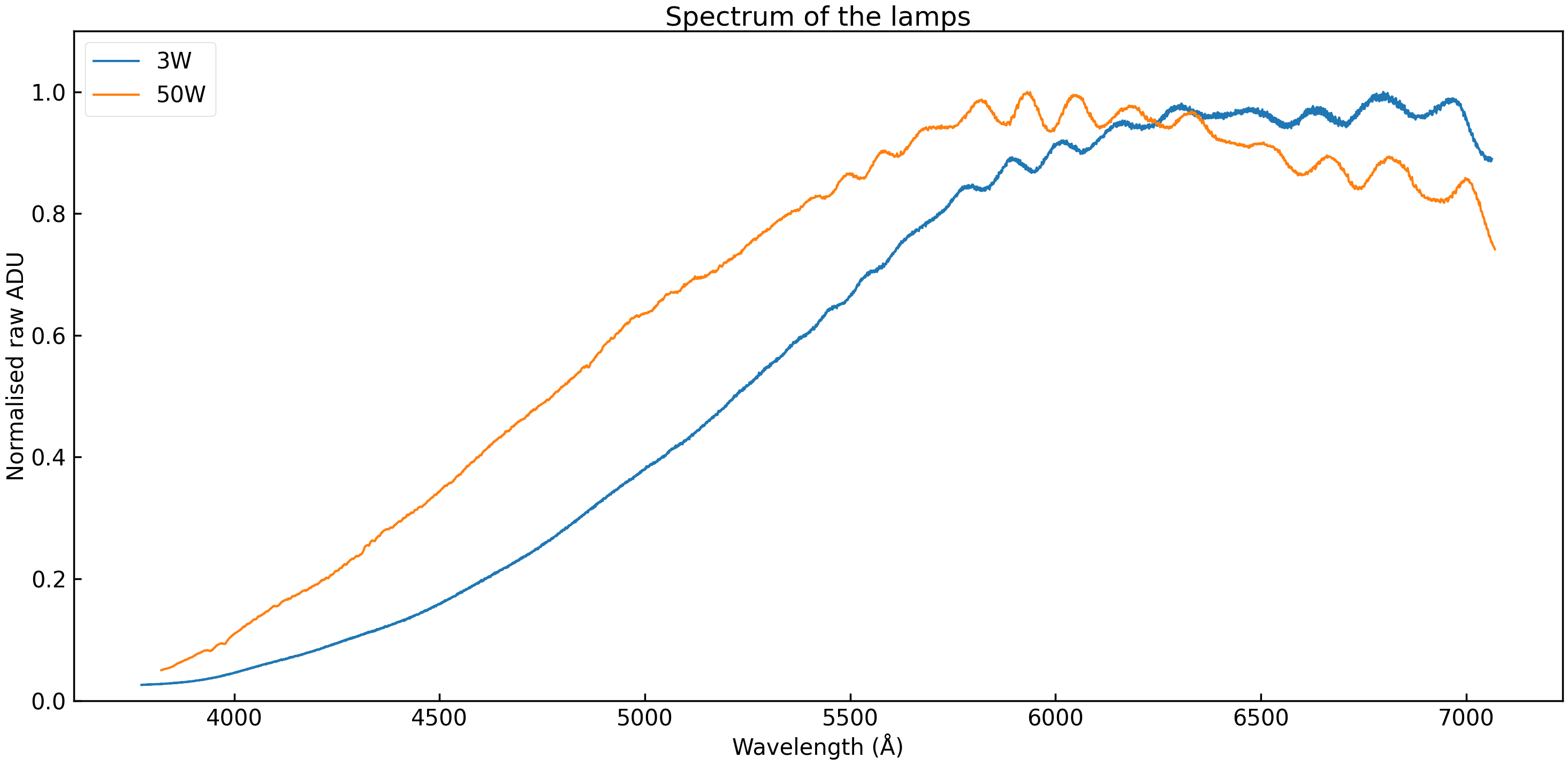}
    \caption{Spectra of 3W and 50W incandescent lamps, showing the peak wavelength shift due to temperature differences.}
    \label{fig:lambSpectrum}
\end{figure}

Based on Wien’s law, we measured the temperatures of two incandescent light bulbs with 12V operating voltage and power ratings of 3W and 50W, respectively. Wien’s law relates the peak emission wavelength $\lambda_{max}$ to the temperature $T$ of the object

\begin{equation}
\label{eqn:LambdaMax}
\lambda_{\text{max}} = \frac{2.898 \times 10^{-3}}{T}
\end{equation}

where \(\lambda_{\text{max}}\) is in meters and \(T\) is in Kelvin. \autoref{fig:lambSpectrum} shows the spectra of the light bulbs, illustrating how peak wavelength shifts as temperature changes.

Using \autoref{eqn:LambdaMax}, we measured the temperature of two light bulbs.

\begin{table}[h]
    \centering
    \caption{Parameters of {\sc SpectrumMate}.}
    \label{comparison}
    \begin{tabular}{cccc}
        \toprule
        Power & Peak wavelength $(\angstrom)$ & Temperature measured ($K$) & Temperature by the manufacturer ($K$)\\
        \midrule
        3 $W$ & 6800 & 4261.76 & 2500\\
        50 $W$ & 5938.46 & 4883.42 & 3350\\
        \bottomrule
    \end{tabular}
\end{table}

The result shows that there is a significant discrepancy between the temperature measured using the peak wavelength and the temperature reported by the manufacturer. This difference can be explained by several factors. First, the peak emission wavelengths observed in the visible range do not correspond to the true blackbody peak for these incandescent light bulbs; the actual emission peaks for the stated temperatures ($2500\ K$ and $3350\ K$) lie in the infrared. Since the spectrograph primarily captures the visible tail of the spectrum, applying Wien’s law to this portion results in an overestimation of the temperature. Second, the calculation assumes the source behaves as an ideal blackbody (with emissivity equal to 1), but real materials, such as tungsten filaments, exhibit wavelength-dependent emissivity that can alter the spectral shape and shift the apparent peak. Finally, instrumental factors such as calibration inaccuracies in the spectrograph’s wavelength scale and sensitivity further contribute to the discrepancy.

\subsection{Filter Transmission Testing}
\begin{figure}[H]
    \centering
    \includegraphics[width=\textwidth]{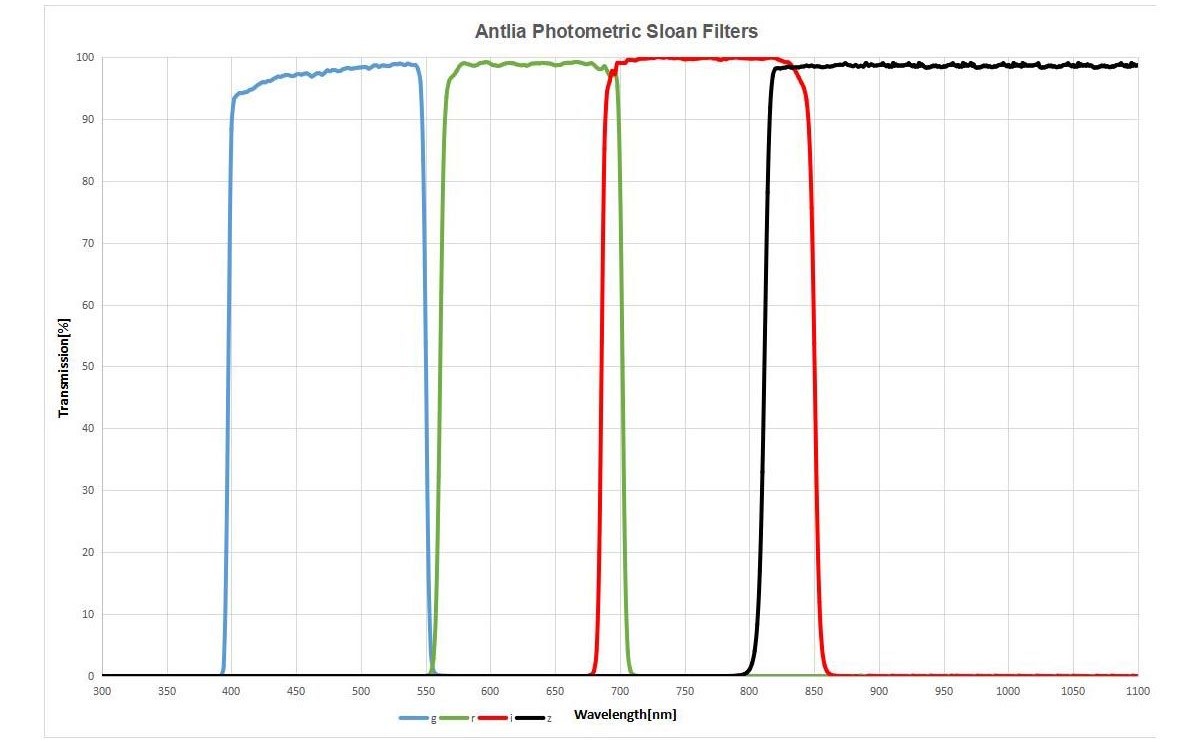}
    \caption{Transmission profile of the filters by the manufacturer. The vertical axis is for transmission (\%), and the horizontal axis is for wavelength ($nm$).}
    \label{fig:ofil-af-griz-curve_1}
\end{figure}

We conducted tests with the {\sc SpectrumMate LR} spectrograph to validate the spectral transmission characteristics of Antlia SLOAN/SDSS (g'r'i'z') Photometric Imaging Filter Set\footnote{\url{https://agenaastro.com/antlia-sloan-sdss-griz-photometric-imaging-filter-set-50mm-round.html?srsltid=AfmBOorLiYmKJ8hMPEmfjryqhooJlT1qK1t82eyp_IUovZ33WQkkLdjv}}, which contains 4 filters, with technical specifications provided by the manufacturer as shown in \autoref{table:4filters} and a transmission profile provided by the manufacturer shown in \autoref{fig:ofil-af-griz-curve_1}.

\begin{table}[h]
    \centering
    \caption{Parameters of {\sc SpectrumMate}.}
    \label{table:4filters}
    \begin{tabular}{lcccc}
        \toprule
        & g'  & r' & i' & z' \\
        \midrule
        Bandpass $(nm)$ & 401-550 & 562-695 & 695-844 & $>$ 820 \\
        Peak Transmission $(\%)$ & 95 & 98 & 98 & 98 \\
        \bottomrule
    \end{tabular}
\end{table}

The transmission tests involved comparing the spectra of a light source in both filtered and unfiltered configurations. A combination of scattered sunlight and incandescent lamps was used as the light source. This hybrid setup was chosen because incandescent lamps lack ultraviolet (UV) radiation, necessitating the inclusion of sunlight to ensure adequate UV coverage. The transmission ratio, defined as the ratio of the filtered spectrum to the unfiltered spectrum, was calculated to accurately quantify the spectral range of each filter. To extend the observed spectral range beyond the standard coverage of the {\sc SpectrumMate LR}, the diffraction grating was slightly tilted to capture wavelengths on both the ultra violet (UV) and near-infrared (NIR) ends of the spectrum. This approach ensured a comprehensive analysis of the filters' transmission properties across a broad wavelength range.

\begin{figure}[H]
    \centering
    \includegraphics[width=\textwidth]{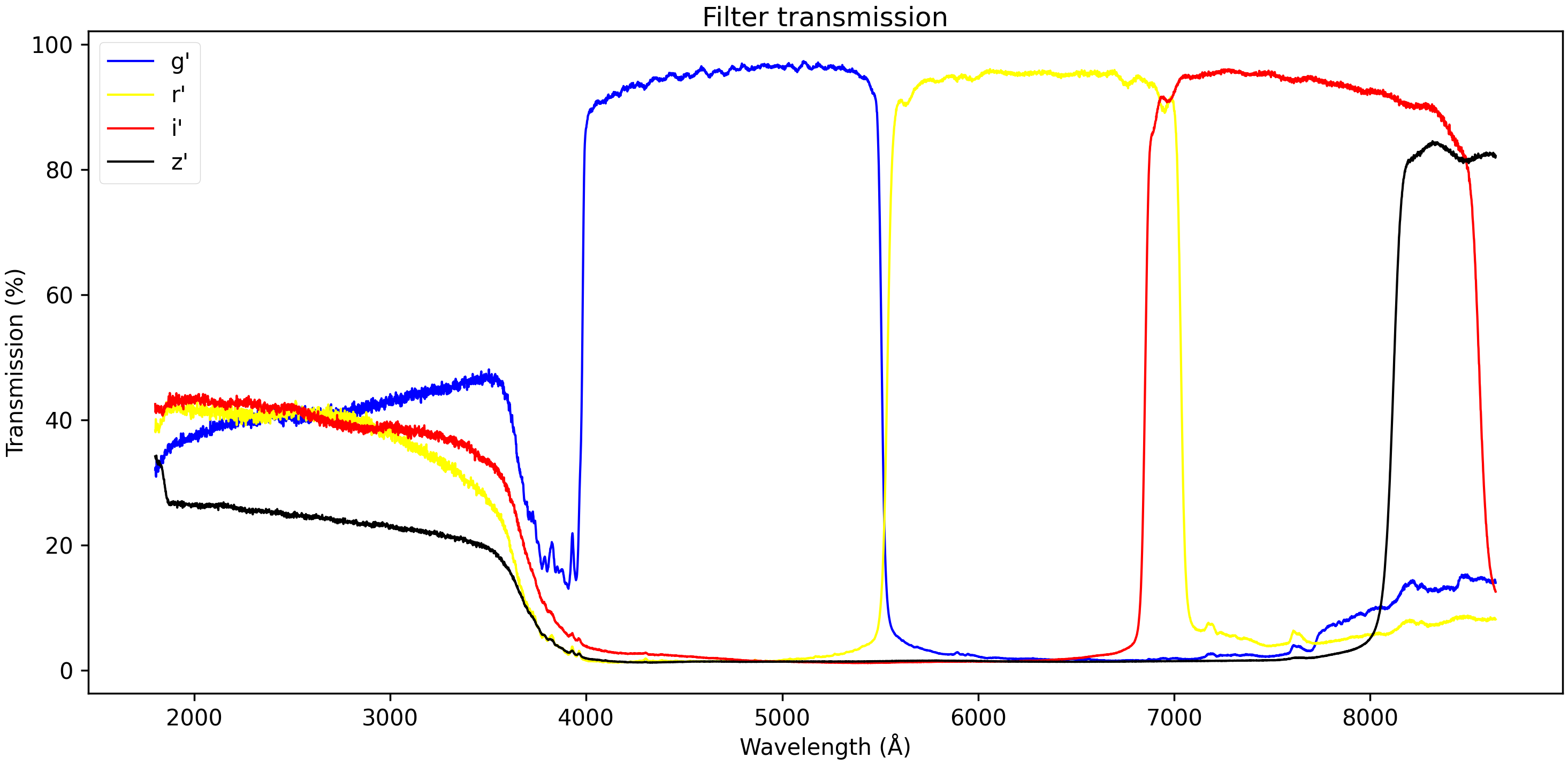}
    \caption{Transmission spectra of the filters measured using \textsc{SpectrumMate LR}.}
    \label{fig:FilterTransmission}
\end{figure}

As shown in \autoref{fig:FilterTransmission}, in the visible range, the filter showed a high transmission percentage (more than 90\%) across each band, with a well-defined cutoff, while its performance decreased significantly in the UV range (lower than 50\%). This decrease can be attributed primarily to the glass material of the lenses used in {\sc SpectrumMate LR}, which absorbs most UV radiation due to the intrinsic absorption properties of standard optical glass (e.g., borosilicate or crown glass). Additionally, shorter UV wavelengths tend to scatter more due to Rayleigh scattering, further reducing the system's signal-to-noise ratio (SNR). Another contributing factor may be the detector's reduced sensitivity in the UV range, as many CCD or CMOS sensors exhibit a natural drop in quantum efficiency at shorter wavelengths unless specifically designed for UV detection. These combined factors lead to a noticeable decline in performance in the UV spectrum.

The data used to create the transmission profile can be found at \url{https://doi.org/10.5281/zenodo.14190404}.

\section{Conclusion}
{\sc SpectrumMate LR} has proven to be a powerful and accessible tool for amateur and educational spectroscopy. It successfully captures detailed spectra for stellar classification, allows temperature measurement of light sources using Wien's law, and enables precise verification of filter transmission properties. These capabilities demonstrate that {\sc SpectrumMate LR} is not only functional but also reliable, providing low-resolution spectral data that can support a range of observational and educational applications. Its performance suggests potential for further use in amateur astronomy and basic research, making it a valuable addition for those interested in spectroscopy.

\section*{Acknowledgement}
We were partly supported by a grant from the Simons Foundation (916424, N.H.) in addition to the enthusiastic support of IFIRSE/ICISE staff. We would also like to thank the QNO for their supports.

\clearpage
\bibliographystyle{apalike}
\bibliography{mybib}

\end{document}